\documentclass[preprint,aps,prc,showpacs,nofootinbib]{revtex4}
\usepackage{epsfig}
\usepackage{graphicx}

\newcommand\ba{\begin{eqnarray}}
\newcommand\ea{\end{eqnarray}}
\newcommand\be{\begin{eqnarray}}
\newcommand\ee{\end{eqnarray}}
\newcommand\nn{\nonumber}

\newcommand{\br}[1]{\left( #1 \right)}
\newcommand{\brs}[1]{\left[ #1 \right]}

\begin{document}

\title{Measuring C-odd correlations at
lepton-proton and photon-proton collisions}

\author{A. I. Ahmadov \footnote{E-mail: Azad.Ahmedov@sunse.jinr.ru}}
\affiliation{JINR, 141980 Dubna, Moscow region, Russian Federation}
\affiliation{Institute of Physics of Azerbaijan National Academy of Sciences, Baku, Azerbaijan}
\author{Yu. M. Bystritskiy \footnote{E-mail: bystr@theor.jinr.ru}}
\affiliation{JINR, 141980 Dubna, Moscow region, Russian Federation}
\author{E. A. Kuraev \footnote{E-mail: kuraev@theor.jinr.ru}}
\affiliation{JINR, 141980 Dubna, Moscow region, Russian Federation}
\author{E. Zemlyanaya}
\affiliation{JINR, 141980 Dubna, Moscow region, Russian Federation}
\author{T. V. Shishkina}
\affiliation{Belarus State University, Minsk, Belarus.}

\date{\today}

\begin{abstract}
We consider the charge-odd correlations (COC) in cross sections of processes
of production of charged particles. The cases of a muonic
pair and pion systems $\pi^+\pi^-$, $\pi^+\pi^-\pi^0$
are considered in detail for electron-proton or photon-proton collisions
in the proton fragmentation region kinematics. COC arise from interference
of amplitudes which describe the different mechanisms of charged
leptons (pions) creation. One of them corresponds to production of
particles in the charge-odd state (one virtual photon
or vector meson annihilation to this system of particles)
and the other corresponds to the charge-even state of produced particles
(creation by two photons).
COC for muon-antimuon pair creation have a pure QED nature and can be
considered as a normalization process. The processes with pion production are
sensitive to some characteristics of proton wave functions and, besides,
can be used for checking the anomalous and normal parts of the effective
pionic lagrangian.

Three electromagnetic currents operator matrix element can be
measured in photon-proton interactions with lepton pair production.
For this aim a charge-odd combination of cross sections
can be constructed as a conversion of leptonic 3-rank tensor
with hadronic ones.
These experiments can be considered as an alternative to deep
virtual Compton scattering.
\end{abstract}

\maketitle

\section{Introduction}
\label{SectionIntroduction}

We want attract an attention to problem of experimental studies
of anomalies of meson lagrangian.
Wess-Zumino-Witten effective meson lagrangian \cite{BKK} (it's anomalous part)
contains four types of anomalies:
$\pi^0 \to 2\gamma$, $\gamma^* \to 3\pi$, $\gamma\gamma \to 3\pi$, $K \bar K \to 3\pi$.
First two of them was experimentally studied up to now
(a lot of information about $\pi^0 \to 2\gamma$ and rather poor one
about $\gamma^* \to 3\pi$). But last two of them wasn't investigated before.
In experiments with fragmentation region kinematics the anomalous vertex
$\gamma\gamma \to 3\pi$ can be measured. That is one of the main purpose of
our paper.

In papers \cite{hpst1,hpst2,hpst3} the asymmetries of two pion
production was investigated with the use of QCD lagrangian based approach.
The main attention there was paid to differential distributions on
azimuthal angles and invariant mass of pion pair produced
in pionization region. This kinematics permits one to investigate
C-odd interference of two and three gluon exchange mechanisms.

In our paper we use alternative approach based on Wess-Zumino-Witten
effective meson lagrangian expressed in terms of hadronic fields. We obtained
distributions on energy fractions of pions in fragmentation region
if initial proton.

\begin{figure}
\begin{center}
\includegraphics[width=10cm]{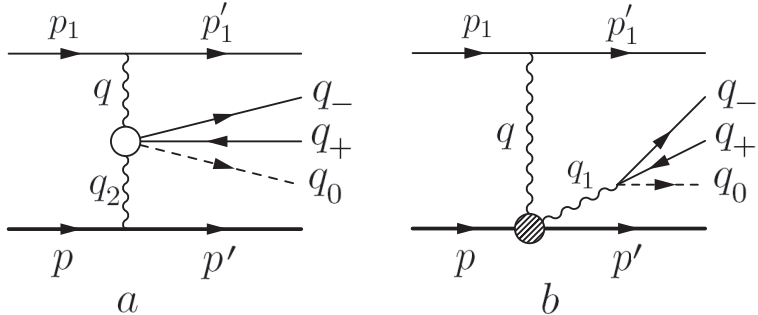}
\caption{The mechanisms of production of muons (pions) pair and
three pions state.}
\label{Fig1}
\end{center}
\end{figure}

The charge-odd contribution to cross section of processes
\ba
e^-(p_1)+p(p)&\to& e^-(p_1') + p(p') + \mu^-(q_-) + \mu^+(q_+), \nn \\
e^-(p_1)+p(p)&\to& e^-(p_1') + p(p') + \pi^-(q_-) + \pi^+(q_+), \label{Reactions}    \\
e^-(p_1)+p(p)&\to& e^-(p_1') + p(p') + \pi^-(q_-) + \pi^+(q_+) + \pi^0(q_0), \nn
\ea
is caused by interference of amplitudes describing the two-photon
mechanism and one photon mechanism of meson set production
(see Fig. \ref{Fig1}, a, b).
The similar quantity can be constructed for inelastic collisions of photon
with hadron with production of lepton pairs. The properties of effective
meson lagrangian as well as three-current correlators matrix elements, averaged
on hadron states can be measured  in relevant experiments, which is a motivation
of our paper.
We consider below the experimental set-up corresponding to
the kinematical region of proton fragmentation,which means
that the recoil proton and mesons created move in
directions close to the initial proton motion in the center of mass system of
initial particles, with invariant mass square of this jet $s_1$ much smaller
than the center of mass square of the total energy of initial particles $s+M^2$
($M$ is proton mass,we put below the electron mass as well as $\mu$ and $\pi$
meson masses $m$ to be small and neglect the terms of order $m^2/s_1$, $s_1/s$).

For laboratory system (initial proton in rest) the angles of emission
of the produced particles can be of order of unity (see discussion below).

This quantity can be measured using the combination of the double
differential cross sections
\ba
\frac{d\sigma^{odd}}{d\Gamma}=\frac{1}{2}[F(q_+,q_-,X)-F(q_-,q_+,X)],
\qquad
\frac{d\sigma}{d\Gamma}=F(q_+,q_-,X),
\ea
where $X$ is the characteristics of other particles, $d\Gamma$ is the
phase volume of final particles.

Our paper is organized as follows.
In section \ref{SectionMesonsProduction}
we calculate charge-odd cross section for processes (\ref{Reactions}).
In section \ref{SectionThreeCurrentCorrelators}
the charge-odd inelastic photon-hadron (proton) scattering is
discussed.
In Conclusion we estimate the order of C-odd contribution, give
spectral distribution and discuss the background effects.

\section{Mesons production}
\label{SectionMesonsProduction}

The remarkable feature of such a kinematics-the relevant contribution
to the cross section do not depend on $s$ in high energy limit.
The corresponding matrix elements are proportional to $s$. This fact can be
explicitly seen using the Gribov's representation of nominator of the
virtual photon Green function in Feynman gauge:
\ba
g_{\mu\nu}=g_{\bot\mu\nu}+\frac{2}{s}[p_{1\mu}\tilde{p}_\nu+p_{1\nu}\tilde{p}_\mu],
\ea
with light-like vectors $p_1$, $\tilde{p}=p-\frac{M^2}{s}p_1$,
$p_1^2=\tilde{p}^2=0$.

Matrix element corresponding to one photon mechanism of $\mu^+\mu^-$ meson pair
production ("bremsstrahlung" ones) have a form
\ba
M_1=\frac{(4\pi\alpha)^2}{q_1^2q_2^2}\frac{2}{s}sN_1 s\bar{u}(p')V^p_\mu u(p) J^m_\mu,
\ea
with $q=p_1-p_1'$, $q_1=q_++q_-$, $J^m_\mu=\bar{u}(q_-)\gamma_\mu v(q_+)$ is the
conversion of virtual photon to muon pair current and
\ba
&&N_1=\frac{1}{s}\bar{u}(p_1')\tilde p_\nu\gamma^\nu u(p_1), \nn\\
&&V^p_\mu=\hat{p}_1\frac{\hat{p}'-\hat{q}+M}{d_1}\gamma_\mu+
\gamma_\mu\frac{\hat{p}+\hat{q}+M}{d}\hat{p}_1, \nn\\
&&d=(p+q)^2-M^2, \qquad d_1=(p'-q)^2-M^2. \nn
\ea
When summing over spin states of electron (initial and the scattered) we have
$ \sum|N_1|^2=2$. As well the averaged on waves of initial and the recoil proton
functions the quantity $V_\mu$ will be finite in the high energy limit.

The two photon mechanism matrix element have a form
\ba
M_2=\frac{(4\pi\alpha)^2}{q^2q_2^2}\frac{2}{s}sN_1\bar{u}(p')\gamma_\lambda u(p)
s I_\lambda,
\qquad q_2=p-p',
\ea
with $I_\lambda=\bar{u}(q_-)V^m_\lambda v(q_+)$-two photon conversion to muon pair
current
\ba
V^m_\lambda=\hat{p}_1\frac{\hat{q}_--\hat{q}}{d_-}\gamma_\lambda+
\gamma_\lambda\frac{-\hat{q}_++\hat{q}}{d_+}\hat{p}_1,
\qquad
d_\pm=(q_\pm -q)^2-m^2.
\ea

The similar expression is valid for creation of pion pair bremsstrahlung
matrix element. It can be obtained from the muon pair one by replacing
$J^m_\mu$ by $J^\pi_\mu=(q_- -q_+)_\mu$.
For pion pair creation by two photon mechanism the replacement $I^m_\lambda \to
I^\pi_\lambda$ must be done in the relevant matrix element for muons
\ba
I^\pi_\lambda=\frac{1}{s}p_1^\nu
\brs{\frac{(2q_--q)_\nu(-2q_++q_2)_\lambda}{d_-}+
\frac{(-2q_++q)_\nu(2q_--q_2)_\lambda}{d_+}-2g_{\nu\lambda}}.
\ea

For the case of three pion production
we must replace the one photon conversion to 2 pions current $J_\mu$ by
$J^{3\pi}_\mu=\frac{1}{4\pi^2f_\pi^3}(\mu q_+ q_- q_0)$, with
$(\mu q_+ q_- q_0)=\epsilon_{\mu\alpha\beta\gamma}q_+^\alpha q_-^\beta q_0^\gamma $.

For two photon conversion to 3 pions we use (see \cite{BKK} for details)
\be
\Pi_\nu&=&\frac{p_1^\nu}{s}[\rho(\mu\nu q q_2)+
(\mu \nu(q_2-q)q_0)-
\nn\\
&&-\frac{q_+^\nu}{q_+q_2}(\mu q q_- q_0)-\frac{q_-^\nu}{q_-q_2}(\mu q q_+ q_0)
-\frac{q_+^\mu}{q_+q}(\nu q_2 q_- q_0)-\frac{q_-^\mu}{q_-q}(\nu q_2 q_+ q_0)],
\ee
with $\rho=\frac{5}{3}-6 (q_+q_-)/(q_++q_-+q_0)^2$. Here $f_\pi=94~\mbox{MeV}$ is the
pion decay constant.

Due to gauge invariance the replacement $p_1 \to q$ in expressions $V_\mu^p,V_\lambda^m,
I^\pi_\lambda,\Pi_\nu$ turns them to zero. Keeping in mind the approximate kinematical
relation $q=\frac{s_1}{s}p_1+q_\bot$, $q_\bot p_1=q_\bot p=0$ with $q_\bot$-transversal
component of the transfer momentum $q$, one can be convinced that all these currents
turns to zero at $q_\bot \to 0$ limit.

At this stage we use the Sudakov's parametrization of
4-momenta of the problem (see Appendix \ref{AppendixSudakovParametrization}).
Accepting it we perform the phase volume
of the process of type $2\to 4$  and $2\to 5$ defined as:
\ba
d\Gamma_4&=&\frac{(2\pi)^4}{(2\pi)^{12}}\frac{d^3p_1'}{2E_1'}\frac{d^3q_+}{2E_+}
\frac{d^3q_-}{2E_-}\frac{d^3p'}{2E'}\delta^4(p_1+p-p_1'-p'-q_+-q_-); \nn\\
d\Gamma_5&=&\frac{(2\pi)^4}{(2\pi)^{15}}\frac{d^3p_1'}{2E_1'}\frac{d^3q_+}{2E_+}
\frac{d^3q_-}{2E_-}\frac{d^3q_0}{2E_0}\frac{d^3p'}{2E'}\delta^4(p_1+p-p_1'-p'
-q_+-q_--q_0) \nn
\ea
to the form
\ba
d\Gamma_4&=&\frac{1}{(2\pi)^8}\frac{d^2q^\bot~d^2q_+^\bot~d^2q_-^\bot~dx_+~dx_-}{8sx_+x_-(1-x_+-x_-)}; \nn \\
d\Gamma_5&=&\frac{1}{(2\pi)^{11}}\frac{d^2q^\bot~d^2q_+^\bot~d^2q_-^\bot~d^2q_0^\bot~dx_+~dx_-~dx_0}{16sx_+x_-x_0(1-x_+-x_--x_0)}. \nn
\ea
Deriving these expressions we had introduced an auxiliary integration
$\int d^4q\delta^4(p_1-p_1'-q)=1$, use the relation $d^3q_i/(2E_i)=d^4q_i\delta(q_i^2-m_i^2)=
\frac{dx_id^2q_{i\bot}}{2x_i}$ with $x_i$-the energy fraction of $i$-th particle
in the center of mass of colliding beams.
Further we denote $q_i^\bot \equiv (0,\vec q_i)$, where
$i=\pm,0$ and $\vec q_i$ is the two-dimensional vector lying in the plane orthogonal
to beam line.

The standard procedure leads to the charge-odd contribution to the cross sections:
\ba
d\sigma^{odd}_{\mu\mu}&=&\frac{16\alpha^4(d^2\vec q/\pi)(d^2\vec q_+/(2\pi))(d^2\vec q_-/(2\pi))dx_+dx_-}
{\pi(q^2)^2q_1^2q_2^2x_+x_-(1-x_+-x_-)}R^{(\mu)}, \nn \\
d\sigma^{odd}_{\pi\pi}&=&\frac{4\alpha^4(d^2\vec q/\pi)(d^2\vec q_+/(2\pi))(d^2\vec q_-/(2\pi))dx_+dx_-}
{\pi(q^2)^2q_1^2q_2^2x_+x_-(1-x_+-x_-)}R^{(\pi)}, \nn \\
d\sigma^{odd}_{3\pi}&=&\frac{4\alpha^4(d^2\vec q/\pi)(d^2\vec q_+/(2\pi))(d^2\vec q_-/(2\pi))(d^2\vec q_0/(2\pi))
dx_+dx_-dx_0}
{\pi(q^2)^2q_1^2q_2^2x_+x_-x_0(1-x_+-x_--x_0)}\br{\frac{M^3}{4\pi f_\pi^3}}^2 R^{(3\pi)}, \nn
\ea
with
\ba
R^{(\mu)}&=&\frac{1}{4}Tr (\hat{p}'+M)V^p_\mu(\hat{p}+M)\gamma_\lambda*
\frac{1}{4}Tr\hat{q}_-V^m_\lambda\hat{q}_+\gamma_\mu, \nn \\
R^{(\pi)}&=&\frac{1}{4}Tr (\hat{p}'+M)V^p_\mu(\hat{p}+M)\gamma_\lambda
(q_--q_+)_\mu I^\pi_\lambda, \nn\\
R^{3\pi}&=&\frac{1}{M^4}\frac{1}{4}Tr (\hat{p}'+M)V^p_\mu(\hat{p}+M)\gamma_\lambda
(\mu q_+ q_- q_0)\Pi_\lambda. \nn
\ea
Explicit forms of $V^p_\mu$, $V^m_\lambda$, $I^\pi_\lambda$, $\Pi_\lambda$
in terms of Sudakov's variables are given above;the ones for the case $\vec{q}^2<s_1$
are given in Appendix \ref{AppendixSudakovParametrization}.
For general case the expressions $R_i$ are complicated. For the realistic case
$-q^2=\vec{q}^2<s_1$ they can be considerably simplified. Really, one can perform the
angular averaging on transfer momentum $\vec{q}$ and omit the terms of higher order
on $\vec{q}^2$ in the nominators.

Performing the integration on transversal momenta of mesons we obtain:
\ba
\frac{d\sigma^{odd}_{\mu\mu}}{dx_+dx_-d\vec{q}^2}
&=&
\frac{\alpha^4}{\pi M^2\vec{q}^2}
F^{\br{2\mu}}\br{x_+,x_-}, \\
F^{\br{2\mu}}\br{x_+,x_-}
&=&
\frac{\br{x_+-x_-}\Delta \br{3-14\Delta+16\Delta^2}}
{6\br{1-\Delta}^3}.
\label{F2mu}
\ea
Note that in paper of one of us \cite{Kuraev:1976ww}
 the similar quantity was considered
for process $e^+e^- \to e^+e^-\mu^+\mu^-$.

\begin{figure}
\begin{center}
\includegraphics[width=10cm]{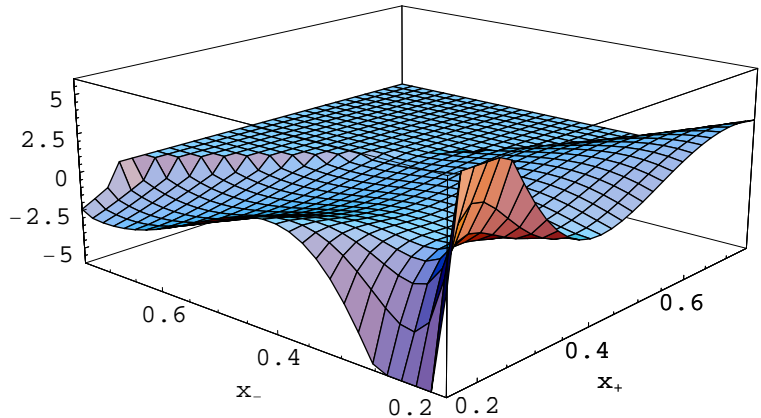}
\caption{Function $100 \cdot F^{\br{2\mu}}\br{x_+,x_-}$ (see (\ref{F2mu}))
dependence of $x_+$ and $x_-$.}
\label{Fig4}
\end{center}
\end{figure}

\begin{figure}
\begin{center}
\includegraphics[width=10cm]{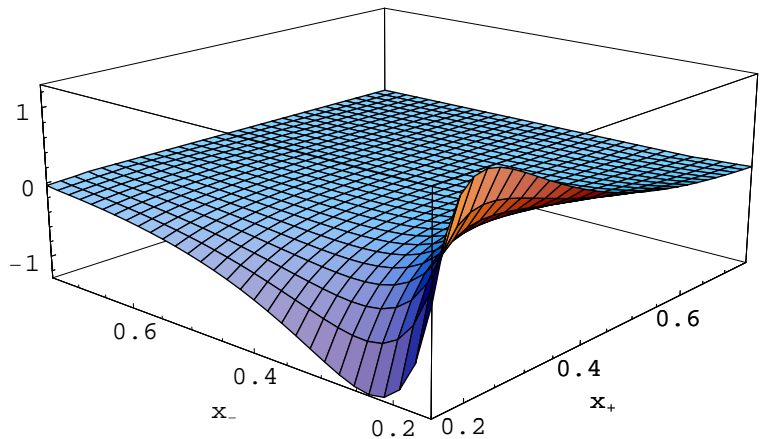}
\caption{Function $F^{\br{2\pi}}\br{x_+,x_-}$ (see (\ref{F2pi}))
dependence of $x_+$ and $x_-$.}
\label{Fig5}
\end{center}
\end{figure}

\begin{figure}
\begin{center}
\includegraphics[width=10cm]{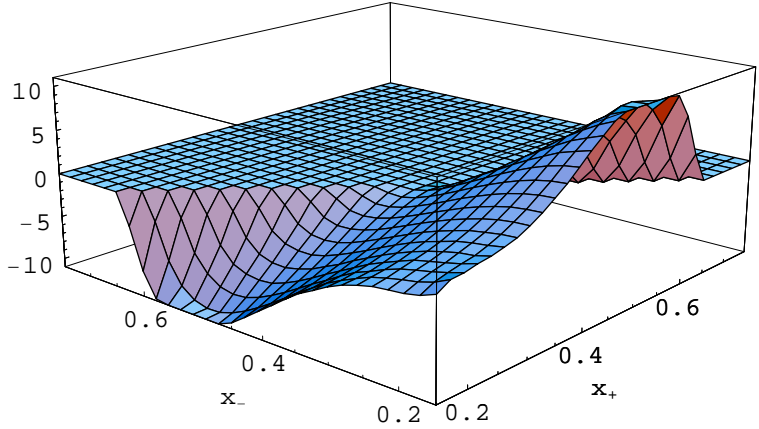}
\caption{Function $100 \cdot F^{\br{3\pi}}\br{x_+,x_-}$ (see (\ref{F3pi}))
dependence of $x_+$ and $x_-$.}
\label{Fig6}
\end{center}
\end{figure}

The similar manipulations for $\pi^+\pi^-$ pair production leads to:
\ba
\frac{d\sigma^{odd}_{\pi\pi}}{dx_+dx_-d\vec{q}^2}
&=&
\frac{\alpha^4}{\pi M^2\vec{q}^2} F^{\br{2\pi}}\br{x_+,x_-}, \\
F^{\br{2\pi}}\br{x_+,x_-}
&=&
\frac{\br{x_+-x_-}\Delta \br{1+\Delta}}{3 \br{1-\Delta}^4}.
\label{F2pi}
\ea
For the case of 3 pions production we have
\ba
\frac{d\sigma^{3\pi}_{odd}}{dx_+dx_-d\vec{q}^2}
&=&
\frac{\alpha^4M^4}{16 \pi^5f_\pi^6 \vec{q}^2}
F^{\br{3\pi}}\br{x_+,x_-},
\\
F^{\br{3\pi}}\br{x_+,x_-}
&=&
\frac{1}{x_+x_-\vec{q}^2}\int\frac{dx_0d^2q_+d^2q_-d^2q_0}
{\pi^3x_0(1-x_+-x_--x_0)q_1^2q_2^2}~\bar R^{3\pi},
\label{F3pi}
\ea
where integration is performed with additional condition
$0.2 < \Delta = 1-x_+-x_--x_0 < 1$ and
$\bar R^{3 \pi}$ is averaged by azimuthal angle of transfer momentum
$\vec q$:
\ba
\bar R^{3\pi} = \int\limits_0^{2\pi} \frac{d\phi}{2\pi}
R^{3\pi},
\ea
where $\phi$ is the azimuthal angle between vector $\vec q$ and the vectors
of problem.
The result of numerical integration for $F^{\br{3\pi}}\br{x_+,x_-}$ as
well as $F^{\br{2\mu}}$ and $F^{\br{2\pi}}$ are given in
the tables \ref{TableF2mu}, \ref{TableF2pi}, \ref{TableF3pi}.
For more illustrative presentation of these functions dependence
see Fig. \ref{Fig4}, \ref{Fig5} and \ref{Fig6}.

\section{Probing three current correlator in charge-odd experimental set-up of
photon-proton collisions}
\label{SectionThreeCurrentCorrelators}

In photon-proton collisions with production of lepton pair
\ba
    \gamma(k) + p(p) \to e^+(q_+) + e^-(q_-) + X(q_h)
\ea
with charge odd experimental set-up
\ba
\frac{d\sigma_{odd}}{d\Gamma}=
\frac{1}{2}\brs{F(q_+,q_-,q_h)-F(q_-,q_+,q_h)},
\qquad
\frac{d\sigma}{d\Gamma}=F(q_+,q_-,q_h),
\ea
where $d\Gamma$ is the phase volume of final particles including
leptons, gives the possibility
to measure the three electromagnetic currents correlator
\ba
H_{\mu\nu\lambda}=<p|J_\mu(k)J_\nu(q)J_\lambda(q_1)|p>.
\ea

\begin{figure}
\begin{center}
\includegraphics[width=10cm]{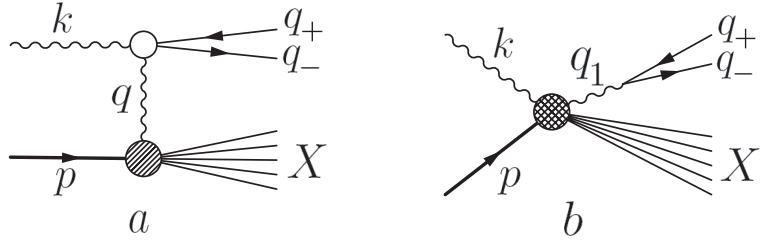}
\caption{The two mechanisms of lepton pair creation:
(a) production of lepton pair in charge-even state,
(b) is the bremsstrahlung mechanism.}
\label{Fig2}
\end{center}
\end{figure}

Really in such kind of experiment the interference of
amplitudes of two mechanisms of lepton pair creation can be measured.
One of them (two-photon ones) corresponds to charge-even state of lepton pair,
another one (bremsstrahlung mechanism) describes the creation of a pair
by the single virtual photon (see Fig. \ref{Fig2}, a, b).
\begin{figure}
\begin{center}
\includegraphics[width=10cm]{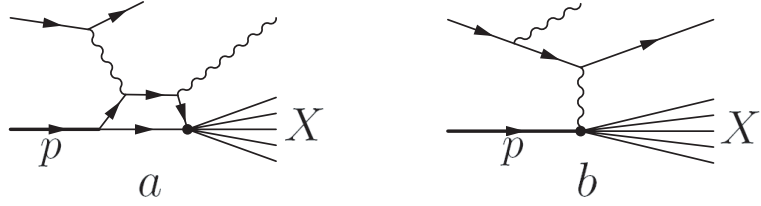}
\caption{DVCS alternative to $\gamma p$
 DIS from Fig. \ref{Fig2}.}
\label{Fig3}
\end{center}
\end{figure}

Charge-odd cross section can be written in form
\ba
\frac{d\sigma}{d\Gamma_+ d\Gamma_-}\sim \frac{\alpha^3}{q^2q_1^2}
L^{(e)}_{\mu\nu\lambda}H_{\mu\nu\lambda} d\Gamma_h,
\ea
with $d\Gamma_h$ is the phase volume of the final hadron system.
Leptonic tensor $L^{(e)}_{\mu\nu\lambda}$, (we neglect lepton mass)
\ba
L^{(e)}_{\mu\nu\lambda}=\frac{1}{4}Tr \left[\hat{q}_-O_{\mu\nu}\hat{q}_+\gamma_\lambda\right],
\ea
with
\ba
O_{\mu\nu}=\frac{1}{\kappa_-}\gamma_\mu(\hat{q}_--\hat{k})\gamma_\nu+
\frac{1}{\kappa_+}\gamma_\nu(-\hat{q}_++\hat{k})\gamma_\mu, \kappa_\pm=2kq_\pm
\ea
obey the gauge conditions $L^{(e)}_{\mu\nu\lambda}k^\mu=L^{(e)}_{\mu\nu\lambda}q^\nu=
L^{(e)}_{\mu\nu\lambda}q_1^\lambda=0$ with
\ba
k^2=q_\pm^2=0, \qquad k+q=q_1=q_++q_-.
\ea
Leptonic tensor can be written in explicitly gauge invariant form:
\ba
L^{(e)}_{\mu\nu\lambda}&=&Q_\lambda T^Q_{\mu\nu}+P_\nu T^P_{\mu\lambda}+R_\mu T^R_{\nu\lambda},
\nn\\
Q_\lambda&=&\frac{1}{2}\br{\frac{1}{\kappa_+}-\frac{1}{\kappa_-}}[\kappa_+\tilde{q}_{-\lambda}+
s_1\tilde{k}_\lambda], \nn\\
P_\nu&=&\frac{s_1}{2}\br{\frac{1}{\kappa_-}-\frac{1}{\kappa_+}}\tilde{k}_\nu+
\frac{\kappa_-+\kappa_+}{2\kappa_-}\tilde{q}_{-\nu}, \nn\\
R_\mu&=&\frac{s_1}{2}\br{\frac{1}{\kappa_-}-\frac{1}{\kappa_+}}\tilde{k}_\mu-
\frac{2s_1+\kappa_++\kappa_-}{2\kappa_-}\tilde{q}_{-\mu}, \nn
\ea
\ba
\tilde{q}_{-\mu}&=&q_{-\mu}-\frac{q_-k}{q_+k}q_{+\mu},                 \qquad \tilde{k}_\mu=k_\mu, \nn\\
\tilde{q}_{-\nu}&=&q_{-\nu}-\frac{q_-q}{q_+q}q_{+\nu},                 \qquad \tilde{k}_\nu=k_\nu-\frac{k q}{q_+q}q_{+\nu}, \nn\\
\tilde{q}_{-\lambda}&=&q_{-\lambda}-\frac{q_-q_1}{q_+q_1}q_{+\lambda}, \qquad \tilde{k}_\lambda=k_\lambda-\frac{k q_1}{q_+q_1}q_{+\lambda}, \nn
\ea
besides
\ba
T^Q_{\mu\nu}=\tilde{g}_{\mu\nu}+\frac{\tilde{k}_\nu\tilde{q}_{-\mu}}{kq};
\quad
T^P_{\mu\lambda}=\tilde{g}_{\mu\lambda}+\frac{\tilde{k}_\lambda\tilde{q}_{-\mu}}{kq_1};
\quad
T^R_{\nu\lambda}=\tilde{g}_{\nu\lambda}+\frac{\tilde{q}_{-\nu}(\tilde{q}_--\tilde{k})_\lambda}{q q_1};
\ea
and, finally
\ba
\tilde{g}_{\mu\nu}=g_{\mu\nu}-\frac{k_\nu q_\mu}{kq};
\quad
\tilde{g}_{\mu\lambda}=g_{\mu\lambda}-\frac{k_\lambda q_{1\mu}}{kq_1};
\quad
\tilde{g}_{\lambda\nu}=g_{\lambda\nu}-\frac{q_\lambda q_{1\mu}}{q q_1}.
\ea
The form of hadronic 3-rank tensor depends on experimental
conditions of detection of hadron jet particles. It won't
be touched here.

\section{Conclusion}

The processes with meson production mentioned above can be studied at
such a facilities as
HERA and HERMES. Photon-hadron interaction processes (the analog of
DIS experiments) can be realized at the facilities with the high energy
photon beams.
The effective meson lagrangian predictions can be examined for 2 and 3 pions
production for the experiments of the first class. In particular the
anomaly $2\gamma\to 3\pi$ can be measured.

In \cite{hpst1,hpst2,hpst3} charge-asymmetry in $2\pi$ production
at virtual photon-proton scattering was investigated in detail. There
was discussed the consequences of Pomeron and Odderon characteristics
measurement.

Our results concern definite process of two and three pion production at
$e-p$ collisions. We obtained some concrete values of asymmetries which is
presented in form of tables and plots convenient for comparison with
corresponding experimental data. This values change from $10^{-4}$ \% to
several percents and presumably can be measured at facilities like CEBAF.

The accuracy of our results is determined by omitted terms and doesn't exceed
10 \%. The uncertainties appear from neglecting the pion final state
interaction (which is small for rather large invariant masses of pion pairs)
and systematical omitting of terms
\ba
    O\left(
        \frac{\alpha}{\pi},
        \frac{\vec q^2}{s},
        \frac{m_\pi^2}{M^2},
        \frac{s_1}{s}
    \right)
    \sim
    O\left(
        0.5 \%,
        3-4 \%,
        2 \%,
        3-5 \%
    \right),
\ea
and the total accuracy is of order 5-7 \%.

For experiments with photon hadron production the three electromagnetic
currents correlations can be studied. Unfortunately these correlations
are very poorly investigated in experiments as well as theoretically
\cite{Ioffe:1982qb}.

Our results was obtained in frames of QED with point-like mesons. In real
applications we must include formfactors of pion $J_\mu^{2\pi}\to F_\pi(q_1^2)(q_+-q_-)_\mu $.
Another modification is replacement of QED coupling constant used for
proton-photon interaction by ones for proton-$\rho$-meson interaction:
$\alpha^4\to \alpha^2(g_{\rho nn}^2/(4\pi))^2$. Besides we must take
into account the resonance character of vector mesons propagators
\ba
\frac{1}{q_1^2}
\to
Re\br{\frac{1}{q_1^2-M_\rho^2+i M_\rho\Gamma_\rho}}
=
\frac{q_1^2 - M_\rho^2}
{\br{q_1^2-M_\rho^2}^2 + M_\rho^2 \Gamma_\rho},
\ea
for $\gamma^* \to \rho \to 2\pi$ and the similar expression
with replacement $M_\rho, \Gamma_\rho \to M_\omega, \Gamma_\omega$
for $\gamma^* \to \omega \to 3\pi$ case.
All these factors was not included in calculation of spectra
given above.

C-odd effects in two pions production in proton fragmentation region
provides besides the possibility to measure the deviation from point
pion approximation used above.
Really the subprocess of two charged pions production at two photons
collision for the case when one of the is real and another is virtual
\ba
\gamma(q)+\gamma^*(q_2) \to \pi(q_-)+\pi(q_+)
\ea
can be described
in terms of three kinematical singularities free amplitudes
\cite{Arbuzov:1997je}:
\ba
T^{\pi_-\pi_+}_{\rho\sigma}&=&a_1 L^{(1)}_{\rho\sigma}+a_2 L^{(2)}_{\rho\sigma}+
a_3 L^{(3)}_{\rho\sigma}, \\
L^{(1)}_{\rho\sigma}&=& (qq_2) g_{\rho\sigma}-q_{2\sigma}q_\rho, \nn\\
L^{(2)}_{\rho\sigma}&=&-(qq_2)Q_\rho Q_\sigma+(q Q)(q_{2\sigma}Q_\rho-q_\rho Q_\sigma)+
(q Q)^2g_{\rho\sigma}, \nn\\
L^{(3)}_{\rho\sigma}&=&(q Q)(q_2^2g_{\rho\sigma}-q_{2\rho}q_{2\sigma})+Q_\sigma((qq_2)q_{2\rho}-q_2^2q_\rho),
\nn
\ea
with $Q=(q_+-q_-)/2$. All three tensor structures are gauge invariant
\ba
L^{(i)}_{\rho\sigma}q^\sigma=L^{(i)}_{\rho\sigma}q_2^\rho=0.
\ea
 The case of
point-like pions corresponds to the choice
\ba
a^{(1)}_0 = \frac{-1}{\chi_++\chi_-}\br{\frac{\chi_+}{\chi_-}+\frac{\chi_-}{\chi_+}},
\qquad
a^{(2)}_0 = \frac{4}{\chi_+\chi_-},
\qquad
a^{(3)}_0 = 0,
\ea
where $2qQ=\chi_+-\chi_-$, $qq_2=\chi_++\chi_-$, $\chi_\pm=qq_\pm$.
We note that in charge-odd experimental set-up differential cross section
contains the linear combination of amplitudes.

Photon-proton deep inelastic interaction
(see Section \ref{SectionThreeCurrentCorrelators}) can be considered as
an alternative to deep inelastic Compton scattering
(see Fig. \ref{Fig3}, a, b) where as well three current correlator
$H_{\mu\nu\lambda}$ can be measured \cite{DVCS}.

\appendix

\section{Sudakov's parametrization}
\label{AppendixSudakovParametrization}

For light lepton-proton scattering
$e(p_1) + p(p) \to e(p_1') + q(q_-) + \bar{q}(q_+) + \cdots + p(p')$
in high energy limit (keeping in mind the experimental requirement of
detecting the final state particles i.e. we must imply the polar angles between their
3-momenta and the beam axes to be sufficiently large) we can consider the  leptons
($e^\pm,\mu^\pm$) as well as pions $\pi^{+,-,0}$ to be massless. Errors caused
by this assumptions is of order
\ba
O\br{\br{\frac{m_\pi}{M}}^2,\frac{m_\pi^2}{s_1},\frac{s_1}{s}},
\ea
with $m_\pi$, $M$ are masses of pion and proton, $s_1$-invariant mass square of
produced particles (excluding the scattered electron), $s=2pp_1 \gg s_1 \sim M^2$.

Introducing the light-like 4-vector $\tilde{p} = p-p_1(M^2/s)$ we use the standard
Sudakov parametrization of 4-momenta of problem:
\ba
&&q_i=x_i\tilde{p}+\beta_i p_1+q_{i\bot}, \qquad a_\bot p_1=a_\bot p=0,
\qquad \tilde{p}^2=p_1^2=0, \nn\\
&& p'=\Delta\tilde{p}+\beta_p p_1+p'_\bot,
\qquad q^2 = q_\bot^2=-\vec{q}^2,
\qquad p=\tilde{p}+\frac{M^2}{s}p_1,
\nn\\
&&q=p_1-p_1'=\alpha_q\tilde{p}+\beta_q p_1+q_\bot.
\ea
We imply $\vec{q}_i$ to be two-dimensional vectors situated in the plane transversal
to the initial electron direction of motion (chosen as a z-axes direction).

Putting on the mass shell conditions $q_i^2=0$, $(p')^2=M^2$, permits to exclude
the "small" coefficients $\beta_i$:
\ba
\beta_i=\frac{\vec{q}_i^2}{sx_i},
\qquad \beta_p=\frac{(\vec{p}'_i)^2+M^2}{s\Delta}.
\ea
Both light-cone component of transfer momentum $q$, $\alpha_q$, $\beta_q$ are
small (of order of $s_1/s$) so we have $q^2\approx-\vec{q}^2$. The conservation law
reads as
\ba
q+p&=&q_++q_-+...+p', \nn\\
1&=&x_++x_-+...+\Delta, \nn\\
\vec{q}&=&\vec{q}_++\vec{q}_-+...+\vec{p}', \nn\\
\beta_q+\frac{M^2}{s}&=&\beta_++\beta_-+...+\beta_p. \nn
\ea
In the center-of mass of initial particles the quantities $x_i$-are the fractions
of energy of the initial proton. Scattering angles of the set of particles,
moving along initial proton direction of motion $\theta_i$ are small quantities
$\theta_i=\frac{2|\vec{q}_i|}{x_i\sqrt{s}}$.

Special attention must be paid for describing the processes in the laboratory frame,
with resting proton. In this frame the light-like 4-vectors are
\ba
\tilde{p}=\frac{M}{2}(1,-1,0,0),
\qquad
p_1=E(1,1,0,0),
\qquad
s=2ME.
\ea
Energies of pions are $E_i=\frac{x_i M}{2}+\frac{\vec{q}_i^2}{2x_iM}$ and the energy
of the scattered proton is $E'=\frac{\Delta M}{2}+\frac{(\vec{p}')^2+M^2}{2M\Delta}$.
The scattering angles of the pions and recoil proton are the quantities of order of
unity
\footnote{First the relations of this type was obtained by
Benaksas and Morrison (see \cite{VK} and references therein).}:
\ba
\sin\theta_i&=&\frac{2x_i M|\vec{q}_i|}{M^2x_i^2+\vec{q}_i^2}, \nn\\
\tan\theta&=&\frac{2M\Delta|\vec{p}'|}{(\vec{p}')^2+M^2(1-\Delta^2)}. \nn
\ea

We put here the expressions of kinematical invariants entering
$R_i$ in terms of Sudakov's variables:
\ba
d_\pm &=& \br{q_\pm - q}^2 - m^2 =
    -\vec q^2 + 2 \vec q_\pm \vec q - s_1 x_\pm, \nn\\
2 q_\pm q_2 &=& 2\vec{q}_\pm\vec{p}'+\frac{\vec{q}_\pm^2\bar{\Delta}}{x_\pm}-
\frac{x_\pm}{\Delta}[(\vec{p}')^2+\bar{\Delta}M^2], \nn\\
2 q_\pm q &=& -2\vec{q}\vec{q}_\pm+s_1x_\pm, \nn\\
2 q_+ q_- &=& \frac{(\vec{q}_-x_+-\vec{q}_+x_-)^2}{x_- x_+}, \nn\\
2 q_\pm q_0 &=& \frac{(\vec{q}_\pm x_0-\vec{q}_0x_\pm)^2}{x_\pm x_0},\nn\\
&&q_2^2=-\frac{1}{\Delta}[(\vec{p}')^2+\bar{\Delta}^2M^2],
\qquad \bar{\Delta}=1-\Delta, \nn
\ea
\be
q_1^2=(q+q_2)^2=-\vec{q}^2+q_2^2+2\vec{q}\vec{p}'+s_1\bar{\Delta}.
\ee

The quantity $s_1$ for $\pi_+,\pi_- p'$ jet have a form
\ba
s_1 &=& -M^2 + \frac{\vec q_+^2}{x_+} + \frac{\vec q_-^2}{x_-} +
\frac{\vec p'^2 + M^2}{\Delta},\vec{q}=\vec{q}_++\vec{q}_-+\vec{p}';
\ea
and for $\pi_+,\pi_-,\pi_0 p'$ jet is
\be
s_1 &=& -M^2 + \frac{\vec q_+^2}{x_+} + \frac{\vec q_-^2}{x_-} +
\frac{\vec{q}_0^2}{x_0}+\frac{\vec p'^2 + M^2}{\Delta},
\vec{q}=\vec{q}_++\vec{q}_-+\vec{q}_0+\vec{p}'.
\ee

The simplified expressions for vertex functions
$V^p_\mu$, $V^m_\lambda$, $I^\pi_\lambda$, $\Pi_\lambda$
(lowest order on $|\vec q|$ expression) are
\ba
V^p_\mu &=&
    \frac{2 \vec p \vec q}{s_{10}^2 \Delta} \gamma_\mu +
    \frac{\gamma_\mu \hat q_\bot \hat p_1}{s s_{10}} +
    \frac{\hat p_1 \hat q_\bot \gamma_\mu}{s s_{10} \Delta}
,
\qquad
\vec p = \vec q_+ + \vec q_-,
\nn\\
V^m_\lambda &=&
    \frac{2 \vec q \vec r}{s_{10}^2 x_+ x_-} \gamma_\lambda +
    \frac{\hat p_1 \hat q_\bot \gamma_\lambda}{s_1 x_-} -
    \frac{\gamma_\lambda \hat q_\bot \hat p_1}{s_1 x_+}
,
\qquad
\vec r = x_- \vec q_+ - x_+ \vec q_-,
\nn\\
I^\pi_\nu &=&
    \frac{1}{s_{10}}
    \br{
        \frac{2 \vec q_- \vec q}{s_{10} x_-} \br{2 q_+ - q_2}_\nu +
        \frac{2 \vec q_+ \vec q}{s_{10} x_+} \br{2 q_- - q_2}_\nu +
        2 q^\bot_\nu
    }
,\nn\\
\Pi^\nu &=&
    \frac{1}{s}\brs{
        \rho \br{p_1 \nu q_\bot q_2} -
        \br{p_1 \nu q_\bot q_0} -
        \frac{q_+^\nu}{q_+ q_2}\br{p_1 q_\bot q_- q_0} -
        \frac{q_-^\nu}{q_- q_2}\br{p_1 q_\bot q_+ q_0}
    }-
\nn\\
&-&
    \frac{1}{s_{10}}\brs{
        \frac{\vec q_+ \vec q}{q_+ q}\br{\nu q_2 q_- q_0} +
        \frac{\vec q_- \vec q}{q_- q}\br{\nu q_2 q_+ q_0}
    }
,\nn
\ea
with $s_{10}=s_1(\vec{q}=0)$.

\begin{table}
\begin{tabular}{|c|c|c|c|c|c|c|c|c|c|c|c|c|}
\hline
$x_- / x_+$ & 0.15 & 0.20 & 0.25 & 0.30 & 0.35 & 0.40 & 0.45 & 0.50 & 0.55 & 0.60 & 0.65 & 0.70\\

\hline
0.15 & 0.000 & 8.338 & 5.625 & 2.112 & 0.000 & -0.676 & -0.370 & 0.446 & 1.399 & 2.222 & 2.734 & 2.821\\
\hline
0.20 & -8.338 & 0.000 & 0.704 & 0.000 & -0.406 & -0.247 & 0.319 & 1.050 & 1.728 & 2.188 & 2.308 & 2.012\\
\hline
0.25 & -5.625 & -0.704 & 0.000 & -0.135 & -0.123 & 0.191 & 0.700 & 1.235 & 1.641 & 1.795 & 1.610 & 0.000\\
\hline
0.30 & -2.112 & 0.000 & 0.135 & 0.000 & 0.064 & 0.350 & 0.741 & 1.094 & 1.282 & 1.207 & 0.000 & \\
\hline
0.35 & 0.000 & 0.406 & 0.123 & -0.064 & 0.000 & 0.247 & 0.547 & 0.769 & 0.805 & 0.569 & & \\
\hline
0.40 & 0.676 & 0.247 & -0.191 & -0.350 & -0.247 & 0.000 & 0.256 & 0.402 & 0.341 & & & \\
\hline
0.45 & 0.370 & -0.319 & -0.700 & -0.741 & -0.547 & -0.256 & 0.000 & 0.114 & & & &\\
\hline
0.50 & -0.446 & -1.050 & -1.235 & -1.094 & -0.769 & -0.402 & -0.114 & & & & &\\
\hline
0.55 & -1.399 & -1.728 & -1.641 & -1.282 & -0.805 & -0.341 & & & & & & \\
\hline
0.60 & -2.222 & -2.188 & -1.795 & -1.207 & -0.569 & & & & & & &\\
\hline
0.65 & -2.734 & -2.308 & -1.610 & 0.000 & & & & & & & &\\
\hline
0.70 & -2.821 & -2.012 & 0.000 & & & & & & & & &\\

\hline
\end{tabular}
\caption{The result of integration for odd part of spectrum
of $2 \mu$ production $100 \cdot F^{\br{2 \mu}}\br{x_+,x_-}$
(see (\ref{F2mu})) for the
case $\vec q^2 < s_1$.}
\label{TableF2mu}
\end{table}

\begin{table}
\begin{tabular}{|c|c|c|c|c|c|c|c|c|c|c|c|c|}
\hline
$x_- / x_+$ & 0.15 & 0.20 & 0.25 & 0.30 & 0.35 & 0.40 & 0.45 & 0.50 & 0.55 & 0.60 & 0.65 & 0.70\\

\hline
0.15 & 0.000 & 1.191 & 1.250 & 1.039 & 0.800 & 0.594 & 0.432 & 0.309 & 0.217 & 0.148 & 0.098 & 0.061\\
\hline
0.20 & -1.191 & 0.000 & 0.346 & 0.400 & 0.357 & 0.288 & 0.221 & 0.162 & 0.115 & 0.078 & 0.050 & 0.028\\
\hline
0.25 & -1.250 & -0.346 & 0.000 & 0.119 & 0.144 & 0.132 & 0.108 & 0.082 & 0.059 & 0.039 & 0.022 & 0.000\\
\hline
0.30 & -1.039 & -0.400 & -0.119 & 0.000 & 0.044 & 0.054 & 0.049 & 0.039 & 0.028 & 0.017 & 0.000 &\\
\hline
0.35 & -0.800 & -0.357 & -0.144 & -0.044 & 0.000 & 0.016 & 0.020 & 0.017 & 0.011 & 0.005 & &\\
\hline
0.40 & -0.594 & -0.288 & -0.132 & -0.054 & -0.016 & 0.000 & 0.006 & 0.006 & 0.003 & & &\\
\hline
0.45 & -0.432 & -0.221 & -0.108 & -0.049 & -0.020 & -0.006 & 0.000 & 0.001 & & & &\\
\hline
0.50 & -0.309 & -0.162 & -0.082 & -0.039 & -0.017 & -0.006 & -0.001 & & & & &\\
\hline
0.55 & -0.217 & -0.115 & -0.059 & -0.028 & -0.011 & -0.003 & & & & & &\\
\hline
0.60 & -0.148 & -0.078 & -0.039 & -0.017 & -0.005 & & & & & & &\\
\hline
0.65 & -0.098 & -0.050 & -0.022 & 0.000 & & & & & & & &\\
\hline
0.70 & -0.061 & -0.028 & 0.000 & & & & & & & & &\\

\hline
\end{tabular}
\caption{The result of integration for odd part of spectrum
of $2 \pi$ production $F^{\br{2 \pi}}\br{x_+,x_-}$
(see (\ref{F2pi})) for the
case $\vec q^2 < s_1$.}
\label{TableF2pi}
\end{table}

\begin{table}
\begin{tabular}{|c|c|c|c|c|c|c|c|c|c|}
\hline
$x_- / x_+$ & 0.15 & 0.20 & 0.25 & 0.30 & 0.35 & 0.40 & 0.45 & 0.50 & 0.55\\

\hline
0.15 & 0.000 & 1.222 & 1.661 & 3.267 & 5.727 & 8.724 & 11.380 & 12.661 & 10.661 \\
\hline
0.20 & -1.037 & 0.000 & 1.114 & 2.826 & 5.088 & 7.261 & 8.551 & 7.544 & \\
\hline
0.25 & -1.736 & -1.090 & 0.000 & 1.635 & 3.548 & 4.988 & 4.883 & & \\
\hline
0.30 & -3.206 & -2.831 & -1.645 & 0.000 & 1.642 & 2.424 & & &\\
\hline
0.35 & -5.779 & -5.031 & -3.509 & -1.628 & 0.000 & & & & \\
\hline
0.40 & -8.718 & -7.219 & -5.003 & -2.399 & & & & & \\
\hline
0.45 & -11.371 & -8.551 & -4.934 & & & & & & \\
\hline
0.50 & -12.675 & -7.543 & & & & & & & \\
\hline
0.55 & -10.789 & & & & & & & & \\

\hline
\end{tabular}
\caption{The result of numerical integration for odd part of spectrum
of $3 \pi$ production
$100\cdot F^{\br{3\pi}}\br{x_+,x_-}$ (see (\ref{F3pi})) for the
case $\vec q^2 < s_1$.}
\label{TableF3pi}
\end{table}

\end{document}